\documentclass[useAMS]{cJAS2e} 
\usepackage[latin1]{inputenc}

\usepackage{amsmath} 

\usepackage{graphicx} 

\usepackage{epsfig}
\usepackage{color}
\usepackage{psfrag}

\usepackage{amsfonts}
\usepackage{amssymb}
\begin{document}


\articletype{SHORT COMMUNICATION}

\title{A statistical testing framework for evaluating the
quality of measurement processes}

\author{Edgard Nyssen$^{\rm a}$$^{\ast}$%
\thanks{$^\ast$Corresponding author.
Email: ehnyssen@etro.vub.ac.be
\vspace{6pt}} %
and Wolfgang Jacquet$^{\rm b}$\\\vspace{6pt}  %
$^{\rm a}${\em{VUB, ETRO Dept., Pleinlaan 2,
B-1050 Brussel, Belgium}};\\%
$^{\rm b}${\em{VUB, EDWE Dept., Pleinlaan 2,
B-1050 Brussel, Belgium}}\\\vspace{6pt}\received{v0.1 received October 2012}}



\maketitle

\begin{abstract}

In this paper in which we address the evaluation of
measurement process quality, we
mainly focus on the evaluation procedure, as far as
it is based on the numerical measurement outcomes.
We challenge the approach where the ``exact'' value of
the observed quantity is compared to the error interval obtained
from the measurements under test and
we propose a procedure where reference measurements
are used as ``gold standard''.
To this purpose,
we designed a specific t-test
procedure for this purpose, explained here.
We also describe and discuss a numerical simulation experiment
demonstrating the behaviour of our procedure.\bigskip

\end{abstract}

\section{Introduction}
\label{sec:intro}

In an experimental context, a number of circumstances require the
evaluation of a factor influencing the quality of measurements like
the apparatus, its calibration, the context of the measurement set-up
and the person(s) performing the measurement.
In this paper, we
address the question of the optimal use of the measurement outcomes
in the evaluation process.
Note that we don't exclude the use of supplementary
quality criteria in this evaluation process, but
we are convinced that
the measurement results contain sufficient (complementary)
information to justify a more thorough study of their use.

We have chosen as example the
evaluation of measurements performed by students in a student lab.
A teaching assistant explains to a group of freshmen how a given
procedure needs to be performed to measure a specific physical quantity.
The students are asked to repeat this procedure a
number of times, to calculate from the measurement data
a mean measurement value $m$, as well as an estimate $s_m$
of the standard deviation of $m$,
and to state in their
report that they consider the observed physical quantity $\mu$
being characterized by an error interval
$m \pm s_m$.

The evaluation of the students
may include the observation by a teacher or a
teaching assistant of the actions by these students
during the measurements,
as well as the assessment of written reports and/or oral tests.
The outcomes of the measurements are a valuable
source of information regarding the performances of the students.
The measurement data are usually considered to be
normally distributed.
Although it may be interesting to study
the effect on the measurement procedure of
a violation of this simplifying assumption, we will
assume its correctness.
In the specific pedagogical setting the exact value $ \mu $ of
the measured quantity is assumed being known.
The fact whether or not this
value is situated in the reported interval,
or a scaled version $m \pm as_m$ thereof, is
used as a criterion to evaluate the quality of execution of the
measurement procedure.
%
%
We will show in this paper that the described type of
approach needs to be challenged from a
statistical point of
view, and we will propose an alternative
for the assessment of the measurement outcomes.


\section{Theoretical considerations}
\label{sec:theorconsid}

%
Observing a physical quantity by executing a measurement procedure
$ n $ times is equivalent to
drawing a random sample
$ \{ x_1 , x_2 , \ldots, x_n \} $
from a population of measurement data, distributed
about an expectation $\mu$
(the actual value of the measured quantity, treated as unknown)
with a standard deviation $\sigma$.
The first steps of error analysis are:
\begin{itemize}
\item the calculation
of an estimate $m$ for $\mu$, which is the arithmetic sample mean,
\item the calculation
of an estimate $s$ for $\sigma$:
\[ s^2 = \frac{\sum_{i=1}^{n} (x_i - m)^2}{n-1}\mbox{,} \]
\item the calculation
of an estimate $s_m$ for the standard deviation $\sigma_m$ of
the mean value, which is $s_m = s/\sqrt{n}$ by virtue of the
root-n law.
\end{itemize}

In the measurement evaluation procedure mentioned in Section \ref{sec:intro},
comparing $ \mu $ with the scaled error interval
$m \pm as_m$ is equivalent to performing a hypothesis test.
The sample under test consists of the $ n $ measurement outcomes,
the test variable is $ t_m = (m - \mu)s_m^{-1} $,
and the null hypothesis $ \mathrm{H}_0 $
is $ \mathrm{E}\{m\} = \mu $.
Indeed, under this hypothesis and the assumption of normality
of the measurement outcomes,
the distribution of
$ t_m $ is known to satisfy
Student's t-distribution with $n-1$ degrees of freedom.
The test consists of verifying whether $ t_m \notin [-a,a] $
and to reject the null hypothesis in this case.
This procedure boils down to a verification whether
$ \mu \notin [m - a s_m , m + a s_m] $
in which case the validity of the measurements is rejected.
The significance level of the test is $ 1 - P_a $, where
$ P_a = \mathrm{P} (t_m \in [-a,a]) $.
One parameter that can be chosen is the sample size $ n $.
If this is relatively high (say, $n \geq 30$), $ s_m $ can
be approximated by $ \sigma_m $ and
$ t_m \approx (m - \mu)\sigma_m^{-1} $ approximately satisfies
a standard normal distribution.
The parameter $ a $ should be sufficiently high to decrease
the significance level. E.g.\ for high $ n $ and $a=1$,
$ P_1 $ is only 0.68 which means that correctly performed
measurements will only be accepted as such with a probability
of 68\%! For $ n < 30 $ this figure is even worse.
However, increasing $ a $ weakens the test.
In many situations, it will be impossible to find a satisfactory
trade-off for the choice of $a$ that avoids the wrongful
rejection of correct measurements and at the
same time yields a criterion to detect bad measurements that
has a sufficient sensitivity (the rightful rejection ratio)
especially when $n$ is small.

However,
our main criticism does not concern
the choice of $a$ (e.g.\ $a=1$), but the fact that
the evaluation criterion is only
dependent on the measurements
under test.
Consequently,
independently of the choice of $a$,
an increased value for $s_m$, which should be interpreted as a decrease
in measurement quality, actually leads to a increased acceptance of the
measurements.

\section{Methods}

The main problem is that
the parameters $\mu$ and $\sigma^2$ characterizing
correctly acquired measurements are generally unknown.
In the present paper, we propose a methodology
using
the outcomes of a reliable reference measurement
as ground truth for a decision on the
validity of the measurements under test.
In our student evaluation example, this could be realised by
letting a skilled teaching assistant or
lab technician repeat the measurement process
-- say, $N$ times.
This leads to unbiased estimates $m_\mathrm{R}$ and
$s_\mathrm{R}^2$ of the operational parameters $\mu$ and $\sigma^2$
respectively
(provided that the unit(s) of measurement are sufficiently refined
for the effect of discretisation to be negligible\cite{ref:Heit89})
-- the suffix $\mathrm{R}$ stands for ``Reference''.
We will denote by $m_\mathrm{T}$ and $s_\mathrm{T}^2$ the sample
mean and variance of the measurements under \underline{t}est
(hence, suffix $\mathrm{T}$).
Before we describe the measurement evaluation procedure, we formulate
the following underlying assumptions:
\begin{itemize}
\item correctly acquired measurement data satisfy the normal distribution
$\mathrm{N}(\mu,\sigma^2)$,
\item the set of reference measurement outcomes are considered as
a representative sample from the population of ``correctly acquired''
measurements, i.e. $m_\mathrm{R}$ and $s_\mathrm{R}^2$, are unbiased
estimates of $\mu$ and $\sigma^2$, respectively.
\end{itemize}
For comparing the reference measurement data and test measurement data,
two criteria are straightforward candidates: the mean value and the sample variance.
In this section, we address both of them.
In the remainder of the text, we mainly concentrate on the former criterion, because
its use is less obvious.

\subsection{Evaluation on the basis of the mean value of the measurement data}
\label{sec:evalmean}

A sound measurement quality assessment procedure only
should wrongfully accuse
a measurement process of yielding bad outcomes at a specific low rate of, e.g.,
one to one hundred on average.
We designed the formula for the test variable of a
hypothesis test
that leads to the definition of an acceptance interval for the
mean value of the measurement data under evaluation.
The design is such that this interval is solely dependent on
the reference measurements and on the chosen operational parameters
$N$ and $n$ (the number of reference measurements and test measurements
respectively).
\newcommand{\mdiff}{(m_\mathrm{R} - m_\mathrm{T})}%
\newcommand{\repetfactor}{\sqrt{\frac{1}{N} + \frac{1}{n}}}%
\newcommand{\repetfactorinv}{\sqrt{\frac{N\cdot{}n}{N+n}}}%
The hypothesis test is based on the following definition of variable $t$:
\begin{equation}
t = s_\mathrm{R}^{-1} \repetfactorinv \mdiff \mbox{.}
\label{eqn:tdefinition}
\end{equation}
We can show (see Appendix \ref{apx:our_t_test}) that $t$ satisfies
Student's t-distribution with $N-1$ degrees of
freedom under the general assumptions formulated earlier and under the
(null) hypothesis that the measurement process under test is correct,
implying that the resulting measurements
satisfy $ \mathrm{N}(\mu,\sigma^2) $.

The acceptance interval can be derived from the relation between
$t_{\alpha,N-1}$ and $\alpha$ in
\begin{equation}
\mathrm{P}(-t_{\alpha,N-1} < t < +t_{\alpha,N-1}) = 1-\alpha \mbox{.}
\label{eqn:tintervalprob} 
\end{equation}
Replacing $t$ by it's expression from Eq.~(\ref{eqn:tdefinition}),
and reformulating the resulting inequalities yields
\[
\mathrm{P}\left(
| m_\mathrm{T} - m_\mathrm{R} |<
t_{\alpha,N-1}
\left(\sqrt{\frac{1}{N} + \frac{1}{n}}\right) s_\mathrm{R}
\right)
= 1-\alpha \mbox{,}
\]
which defines as acceptance interval for a given $\alpha$:
\begin{equation}
\label{eqn:acceptintv} 
m_\mathrm{R} \pm t_{\alpha,N-1}
\left( \sqrt{\frac{1}{N} + \frac{1}{n}} \right)
s_\mathrm{R}\mbox{.}
\end{equation}
The critical $t$-value $t_{\alpha,N-1}$ can be found from the cumulative
probability function of $t$ for $N-1$ degrees of freedom - considering that
the probability density function is symmetric and consequently
Eq.~(\ref{eqn:tintervalprob}) is equivalent to:
\[
\mathrm{P}(t < t_{\alpha,N-1}) = 1-\alpha/2 \mbox{.}
\]
For a given $\alpha$, one can find $t_{\alpha,N-1}$ in a table of
Student's t distribution -- see, e.g., Ref.~\cite{ref:CRC03}.

The reader may wonder why we don't 
propose one of the
classical t-tests for testing the difference of mean values
for independent samples.
There exist two variants of these tests: one in which the variances
of the data in the two samples
are assumed equal and one where this assumption is not required.
It is obvious that in general the former model does not hold for the case
where test measurements need being compared to reference
measurements.

\newcommand{\sigsqRN}{\frac{s_\mathrm{R}^2}{N}}
\newcommand{\sigsqTn}{\frac{s_\mathrm{T}^2}{n}}
A common formal expression for the second model,
is given by:
\begin{equation}
t = \frac{\mdiff}{\sqrt{\sigsqRN+\sigsqTn}}
\mbox{\ \ where\ \ }
df = \left\lceil 
\frac{\left( \sigsqRN + \sigsqTn \right)^2}{\frac{\left( \sigsqRN \right)^2}{N-1} +\frac{\left( \sigsqTn \right)^2}{n-1}}
\right\rceil \mbox{.}
\label{eqn:classic_t_test}
\end{equation}
The specific equation for $df$ is known as the
Welch-Satterthwaite equation.\cite{ref:netj90}
The model is used in studies where the variances of the underlying
variable $x$ are allowed to be different in the populations
from which the two samples are observed.
We already announce here that we dismiss this model as a basis for
the evaluation of measurements and refer
to sections \ref{sec:numexperiments}
and \ref{sec:discussion} for more details about our reasons
to do so.

\subsection{Evaluation on the basis of the variance of the measurement data}
\label{sec:evalvar}

Here, we can directly derive the acceptance interval for the variance
from a standard F-test.
Under the null hypothesis that $s_\mathrm{R}$ and $s_\mathrm{T}$
have been calculated from correct measurement data -- i.e. two independent
samples of data satisfying the same normal distribution $\mathrm{N}(\mu,\sigma^2)$,
$F=s_\mathrm{T}/s_\mathrm{R}$ is distributed according to
the F-distribution $\mathrm{F}(n-1,N-1)$.

Assuming that reducing the quality of the measurements will increase
the variance of their outcomes, one readily can formulate the acceptance
interval as
\[
[0\ ,\ s_\mathrm{R}\,F_{\alpha,n-1,N-1}] \mbox{,}
\]
where the critical $F$-value $ F_{\alpha,n-1,N-1} $ is derived from
the cumulative probability function:
\[
\mathrm{P}(F < F_{\alpha,n-1,N-1}) = 1-\alpha \mbox{.}
\]
Sometimes, the nature of the measurement process requires considering
the possibility that the reduction of the measurement quality
either increases or decreases the variance of the outcomes --
e.g.\ when the person performing the measurements
systematically tends to round the
observed quantities to the same value.
In this case the acceptance interval should be:
\[
[s_\mathrm{R}\,F_{\frac{\alpha}{2},N-1,n-1}^{-1}\ ,\ %
s_\mathrm{R}\,F_{\frac{\alpha}{2},n-1,N-1}] \mbox{.}
\]

\section{Numerical Experiments}
\label{sec:numexperiments}

We performed some numerical experiments to verify in practice the
theoretical considerations.
Each experiment aimed at estimating the rejection ratio $ \rho $
of measurements satisfying one combination of
$\mu_\mathrm{T}$ and $\sigma_\mathrm{T}$ parameter values by repeating
the simulation of one measurement experiment a number of times with
these specific parameter values.
For the estimation of $ \rho $, we calculated the fraction
$\hat{\rho}_\mathrm{pt}$ of simulations where the outcomes of the
(simulated) measurement experiment are rejected (i.e.\ a point estimate),
as well as an interval estimate of $ \rho $ --
$ [\hat{\rho}_\mathrm{lo},\hat{\rho}_\mathrm{hi}] $ --
at a 95\% confidence interval.
Evidently, for measurements from a correct measurement process
($\mu_\mathrm{T} = \mu$ and $\sigma_\mathrm{T} = \sigma$),
we expect that $ \rho (= \mathrm{E}\{\hat{\rho}_\mathrm{pt}\}) = \alpha $.
We considered as hypothetical measurement experiment a titration performed
by freshmen where the
volume of titrant, necessary to neutralize a standardized quantity of
acid would be $ \mu = 21.35\,\mbox{cm}^3 $.
The reference measurements are produced by a laboratory technician
performing a titration, repeated $ N = 10 $ times,
with an accuracy characterized by
$ \sigma = 0.01\,\mbox{cm}^3 $.
The mean result obtained by the technician is compared to the mean result
obtained by a student who
repeats the titration $ n = 3 $ times and measures according to
parameters $\mu_\mathrm{T}$ and $\sigma_\mathrm{T}$.
In case $\mu_\mathrm{T} = \mu$ and $\sigma_\mathrm{T} = \sigma$, we are
dealing with correctly performed measurements. If in that case
the student's mean titration volume falls outside the
acceptance interval, given
by Eq.~(\ref{eqn:acceptintv}), we are confronted with a wrongful rejection.
In order to perform an accurate estimation of the (rightful or wrongful)
rejection rate $\rho$ for different combinations of $ \mu_\mathrm{T} $,
$ \sigma_\mathrm{T} $, and $\alpha$, we simulated $ 10^6 $
independent experiments, 
where each time the student's outcome is tested using the aforementioned
acceptance interval.
The outcomes of our numerical experiments are summarized
in Table \ref{tab:simnewcrit}.
\begin{table}[th]
\begin{center}
\begin{tabular}{|c|c|c|c|c|c|}
\hline $\mu_\mathrm{T}$ & $\sigma_\mathrm{T}$ &
$\alpha$ & $\hat{\rho}_\mathrm{pt}$ &
$\hat{\rho}_\mathrm{lo}$ & $\hat{\rho}_\mathrm{hi}$ \\
\hline
\hline 21.35 & 0.01 & 0.001 & 0.00098 & 0.00092 & 0.00105 \\
\hline 21.35 & 0.01 & 0.010 & 0.01015 & 0.00995 & 0.01035 \\
\hline 21.35 & 0.01 & 0.050 & 0.04997 & 0.04954 & 0.05039 \\
\hline 21.37 & 0.01 & 0.001 & 0.13939 & 0.13872 & 0.14008 \\
\hline 21.37 & 0.01 & 0.010 & 0.46495 & 0.46397 & 0.46592 \\
\hline 21.37 & 0.01 & 0.050 & 0.77146 & 0.77064 & 0.77229 \\
\hline 21.35 & 0.02 & 0.001 & 0.02723 & 0.02691 & 0.02755 \\
\hline 21.35 & 0.02 & 0.010 & 0.10803 & 0.10742 & 0.10864 \\
\hline 21.35 & 0.02 & 0.050 & 0.24466 & 0.24382 & 0.24550 \\
\hline 
\end{tabular} 
\end{center}
\caption{Experiments simulating measurements
by freshmen compared to a skilled lab technician
($\mu_\mathrm{R} = 21.35$, $\sigma_\mathrm{R} = 0.01$, $N=10$, $n=3$) -- results from $ 10^6 $ simulations are a point estimate
for $\rho$ -- $\hat{\rho}_\mathrm{pt}$ --
and an interval estimate for it (confid.\ level 95\%)
$[\hat{\rho}_\mathrm{lo} , \hat{\rho}_\mathrm{hi}]$ for an
acceptance criterion given by Eq.~(\ref{eqn:acceptintv}).
\label{tab:simnewcrit}}
\end{table}
In order to compare the criterion based on a scaled
error interval to the one presented here, in terms of sensitivity,
we repeated some of the experiments with the same parameters and
calculating the rejection ratio using the former criterion.
The outcomes of these numerical experiments are summarized
in Table \ref{tab:simerrcrit}.
\begin{table}[th]
\begin{center}
\begin{tabular}{|c|c|c|c|c|c|}
\hline $\mu_\mathrm{T}$ & $\sigma_\mathrm{T}$ &
$\alpha$ & $\hat{\rho}_\mathrm{pt}$ &
$\hat{\rho}_\mathrm{lo}$ & $\hat{\rho}_\mathrm{hi}$ \\
\hline
\hline 21.35 & 0.01 & 0.010 & 0.00989 & 0.00970 & 0.01009 \\
\hline 21.37 & 0.01 & 0.010 & 0.12160 & 0.12096 & 0.12224 \\
\hline 21.35 & 0.02 & 0.010 & 0.01000 & 0.00981 & 0.01020 \\
\hline 
\end{tabular}
\end{center}
\caption{Experiments simulating measurements
by freshmen and evaluated on the basis of expression
$ \mu \in m \pm as_m $ as acceptance criterion,
where $a$ corresponds to
$ P_a = 1 - \alpha = \mathrm{P} (t_m \in [-a,a])$
as explained in section \ref{sec:theorconsid}
-- results from $ 10^6 $ simulations are
a point estimate
for $\rho$ -- $\hat{\rho}_\mathrm{pt}$ --
and an interval estimate
$[\hat{\rho}_\mathrm{lo} , \hat{\rho}_\mathrm{hi}]$.
for it (confid.\ level 95\%).
\label{tab:simerrcrit}}
\end{table}
A selection of the numerical experiments, reported by Table \ref{tab:simnewcrit}
(those for $\mu_\mathrm{T} = 21.35$, $\sigma_\mathrm{T} = 0.01$),
have been repeated, but with
the criterion based on Eq.~(\ref{eqn:classic_t_test}).
The outcomes of these experiments are summarized
in Table \ref{tab:simclascrit}.
\begin{table}[ht]
\begin{center}
\begin{tabular}{|c|c|c|c|}
\hline $\alpha$ & $\hat{\rho}_\mathrm{pt}$ &
$\hat{\rho}_\mathrm{lo}$ & $\hat{\rho}_\mathrm{hi}$ \\
\hline
\hline 0.001 & 0.003770 & 0.003652 & 0.003892 \\
\hline 0.010 & 0.019548 & 0.019279 & 0.019821 \\
\hline 0.050 & 0.065122 & 0.064640 & 0.065607 \\
\hline
\end{tabular} 
\end{center}
\caption{Experiments simulating measurements -- same
parameters as for Table \ref{tab:simnewcrit} and
satisfying $\mathrm{H}_0$,
i.e.\ $\mu_\mathrm{T} = 21.35$, $\sigma_\mathrm{T} = 0.01$, but
with as acceptance/rejection criterion the t-test for the model given by
Eq.~(\ref{eqn:classic_t_test}).
\label{tab:simclascrit}}
\end{table}

\section{Discussion}
\label{sec:discussion}


In section \ref{sec:evalmean}, we propose an method for
the evaluation of a measurement process,
based on a non-standard Student's t-test, justified theoretically by
Appendix \ref{apx:our_t_test} and validated by the numerical experiments
described by section \ref{sec:numexperiments}.
Our numerical experiments compare the sensitivity of
this methodology with the ``classical'' approach based on a
(scaled) error interval. In the latter approach
the value of $a$ is chosen to yield an equivalent criterion as
ours in terms of $\alpha$ -- the wrongful rejection ratio.
The experiments demonstrate that with an appropriate choice
of the parameter $N$, the sensitivity of our approach is much higher
than for the (scaled) error interval approach (Tables
\ref{tab:simnewcrit} and \ref{tab:simerrcrit}).
Moreover, note
also that where a bad measurement procedure only affects $\sigma_\mathrm{T}$,
(last row of Table \ref{tab:simerrcrit}), the sensitivity of
the error interval criterion even doesn't exceed $\alpha$, whereas ours
is sensitive both to bias and high $\sigma_\mathrm{T}$.

The reader could wonder why not using one of the
two classical t-tests for independent samples.
We already briefly introduced this question at the
end of subsection \ref{sec:evalmean} and mentioned
the hypothesis test of the difference of mean values under
the assumption that the variances of the data in the
two involved samples are allowed to be different. This
problem is known as the Behrens-Fisher
problem.
In Ref.~\cite{ref:ross87}, it is already pointed out that for this problem
``There is no completely satisfactory solution known''.
One very popular solution, found in most text books,
is the difference of means test for unequal population variances,
using the Welch-Satterthwaite equation --
see Eq.~(\ref{eqn:classic_t_test}).\cite{ref:Welch38}
One should realize that $t$ in this equation is generally
only approximately distributed according
to Student's t-distribution.
The shortcomings of the technique
when both $N$ and $n$ are small, with nevertheless an important discrepancy,
are demonstrated by our numerical simulation
experiments (Table \ref{tab:simclascrit}), where
under the null hypothesis
the wrongful rejection ratio is systematically significantly larger
than the chosen significance level $\alpha$.
Moreover, the sample variances
$ s_\mathrm{R} $ and $ s_\mathrm{T} $ are treated as equivalent in
the calculation of the estimate of the variance of the denominator
of the expression for $t$.
In our case, we assume that $ s_\mathrm{R}^2 $ is an
unbiased estimator
for $ \sigma^2 $, a parameter that (together with the constants
$N$ and $n$) fully determines the distribution of $\mdiff$ under the
null hypothesis.
It is reasonable to assume that in most cases
the variance of the data from badly performed
measurements is greater than the variance of the outcomes
of correctly performed measurements.
In that case, treating the sample variances as equivalent
would boil down to weakening the test, especially when considering that
normally $n < N$,
causing the term in $s_\mathrm{T}^2$
to dominate the term in $s_\mathrm{R}^2$ in the denominator of
the expression for $t$.
In the test, proposed here,
we decided not to incorporate
$s_\mathrm{T}^2$ in the expression for $t$ -- see
Eq.~(\ref{eqn:tdefinition}) --
and to retain only the ``reference''
sample standard deviation $s_\mathrm{R}$.
An additional advantage is that this approach yields an acceptance
interval that only needs to be calculated once for the evaluation of
measurement outcomes from several students, since this interval
is only dependent on the reference measurements by the
laboratory technician.
The numerical simulation experiments
demonstrate that for correct measurement processes
(Table \ref{tab:simnewcrit} for
$\mu_\mathrm{T} = 21.35$ and $\sigma_\mathrm{T} = 0.01$),
the wrongful rejection ratio is
consistent with the chosen significance level
and therefore perfectly controllable.
The effect on $\rho$ of a bias in the
measurements under evaluation is demonstrated with $\mu_\mathrm{T} = 21.37$
and shows how this $\rho$ (the -- this time rightful -- rejection rate)
increases at the expense of an increasing $\alpha$.
This means for our specific simulated measurements
example that if we find such bias of $0.02$ sufficiently high
for a measurement process to be qualified as bad, and we are satisfied with
identifying 77\% of the measurement processes affected by such bias,
we must accept to wrongfully reject 5\% of the correct
measurement processes.
One may conceive to
introduce in the procedure additional information about measurement quality
by combining the test on the basis
of the mean value of the measurements (subsection \ref{sec:evalmean}) with
the one based on the variance (subsection \ref{sec:evalvar}), but
then one should take care to reduce the $ \alpha $ of each test by
one half to bring the significance level of the combined
test to (at most) $ \alpha $ (Bonferroni correction).

\bibliographystyle{cJAS}
\bibliography{Statistics_papers}

\newcommand{\bibfont}{\fontsize{10}{12}\selectfont} \newcommand{\noopsort}[1]{}
  \newcommand{\printfirst}[2]{#1} \newcommand{\singleletter}[1]{#1}
  \newcommand{\switchargs}[2]{#2#1}
\begin{thebibliography}{1}
\providecommand{\url}[1]{\texttt{#1}}
\providecommand{\urlprefix}{URL }

\bibitem{ref:Heit89}
D.F. Heitjan, \emph{Inference from grouped continuous data: A review}, Stat.
  Sci. 4 (1989), pp. 164--179.

\bibitem{ref:netj90}
J. Neter, W. Wasserman, and M.H. Kutner, \emph{Applied linear statistical
  models : regression, analysis of variance, and experimental designs},
  {Homewood, Ill. : Irwin} (1990).

\bibitem{ref:ross87}
S.M. Ross, \emph{Introduction to probability and statistics for engineers and
  scientists}, {John Wiley \& Sons, Inc} (1987).

\bibitem{ref:Welch38}
B.L. Welch, \emph{The significance of the difference between two means when the
  population variances are unequal}, Biometrika 29 (1938), pp. 350--362.

\bibitem{ref:CRC03}
D. Zwillinger (ed.), \emph{{CRC} -- {S}tandard mathematical tables and
  formul{\ae}}, 31st ed., {Chapman \& Hall/CRC} (2003).

\end{thebibliography}

\appendices

\section{Student evaluation based on the mean measurement value --
theoretical background}
\label{apx:our_t_test}
The goal of this appendix is to demonstrate that the expression
for $t$ in Eq.~(\ref{eqn:tdefinition})
satisfies Student's t-distribution
with $N-1$ degrees of freedom.

Let us start from the formal definition of Student's t-distributed
variable, consisting of the following
elements:\cite{ref:CRC03}
\begin{enumerate}
\item[(D1)] If the random variable $X$ is normally distributed with mean
0 and variance $\sigma^2$, and \ldots
\item[(D2)] \ldots if $Y^2 / \sigma^2$ has a
$\chi^2$ distribution with $\nu$ degrees of freedom, and \ldots
\item[(D3)] \ldots if $X$ and $Y$ are independent, \ldots
\item[(D4)] \ldots then $ t = \frac{X \sqrt{\nu}}{Y} $ is distributed as a
t-distribution with $\nu$ degrees of freedom.
\end{enumerate}

As linear combination of normally distributed terms $m_\mathrm{R}$
and $m_\mathrm{T}$, expression $\mdiff$ satisfies a normal distribution.
Since $m_\mathrm{R}$ and $m_\mathrm{T}$ have the same expectation $\mu$,
the expectation of $\mdiff$ is zero. Also, we are dealing with the
mean values of independent sets of measurement data.
Therefore
$m_\mathrm{R}$ and $m_\mathrm{T}$ are statistically independent,
resulting in the variance of $\mdiff$ being the sum of
the variances of $m_\mathrm{R}$ and $m_\mathrm{T}$.
Formally:
\[ 
\mdiff :
\mathrm{N} \left( 0,  \sigma^2 \left( \frac{1}{N} + \frac{1}{n} \right) \right)\mbox{.}
\]
This allows us to introduce a variable $X$ and
to equate it to an expression that satisfies
element (D1) of the definition
of the t-distribution, formulated earlier:
\begin{equation}
X = \repetfactorinv \mdiff\ \ :\ \ \mathrm{N}(0,\sigma^2)\mbox{.}
\label{eqn:Xdef}
\end{equation}

Let us now introduce a variable
\begin{equation}
Y = \sqrt{\nu}\, s_\mathrm{R} \mbox{\ \ \ with\ \ \ } \nu = N - 1\mbox{.}
\label{eqn:Yandnudef}
\end{equation}
From this definition, and the fundamental properties of the sample
variance of a normally distributed variable
follows that $Y^2/\sigma^2$ has a $\chi^2$ distribution with $\nu$
degrees of freedom, which satisfies element (D2) of the aforementioned
definition.

Finally note that $X$ and $Y$ are independent -- definition element
(D3) -- since on the one hand
$ m_\mathrm{R} $ and $ s_\mathrm{R} $ are mutually independent as mean and variance of
data of the same sample and on the other hand, $ m_\mathrm{T} $ and $ s_\mathrm{R} $
are independent as statistics of two independent sets of sample data.

This means that
$ t = \frac{X \sqrt{\nu}}{Y} $ satisfies the t-distribution according to
definition element (D4), but substituting $X$, $Y$ and $\nu$ in the latter
expression for $ t $, using Eqs.~(\ref{eqn:Xdef}) and (\ref{eqn:Yandnudef})
yields exactly Eq.~(\ref{eqn:tdefinition}).

\end{document}